\documentclass[aps,showpacs,twocolumn,10pt,prl]{revtex4-1}

\usepackage{amsmath}
\usepackage{amsthm}
\usepackage{amsfonts}
\usepackage{amssymb}
\usepackage{graphicx}

\begin{document}
\title{Quantum-classical correspondence in circularly polarized high harmonic generation}

\author{F. Mauger}
\author{A.~D. Bandrauk}
\affiliation{
Laboratoire de Chimie Th\'eorique, Facult\'e des Sciences, Universit\'e de Sherbrooke, Sherbrooke, Qu\'ebec, Canada J1K 2R1}

\author{A. Kamor}
\affiliation{School of Physics, Georgia Institute of Technology, Atlanta, GA 30332-0430, USA}
\affiliation{Centre de Physique Th\'eorique, CNRS -- Aix-Marseille Universit\'e, Campus de Luminy, case 907, F-13288 Marseille cedex 09, France}

\author{T. Uzer}
\affiliation{School of Physics, Georgia Institute of Technology, Atlanta, GA 30332-0430, USA}

\author{C. Chandre}
\affiliation{Centre de Physique Th\'eorique, CNRS -- Aix-Marseille Universit\'e, Campus de Luminy, case 907, F-13288 Marseille cedex 09, France}


\begin{abstract}
Using numerical simulations, we show that atomic high order harmonic generation, HHG, with a circularly polarized laser field offers an ideal framework for quantum-classical correspondence in strong field physics. With an appropriate initialization of the system, corresponding to a superposition of ground and excited state(s), simulated HHG spectra display a narrow strip of strong harmonic radiation preceded by a gap of missing harmonics in the lower part of the spectrum. In specific regions of the spectra, HHG tends to lock to circularly polarized harmonic emission. All these properties are shown to be closely related to a set of key classical periodic orbits that organize the recollision dynamics in an intense, circularly polarized field. 
\end{abstract}
\maketitle

\begin{figure}
	\centering
	\includegraphics[width=\linewidth]{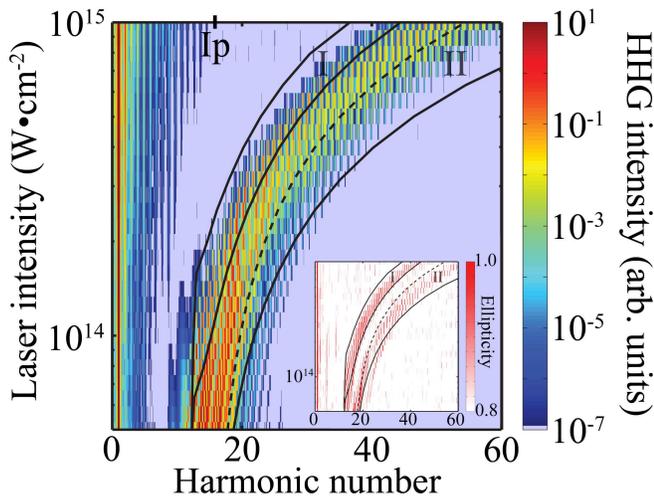}
	\caption{\label{fig:HHG_spectrum} 
	(color online) HHG spectrum (main panel) obtained from numerical integration of Eq.~(\ref{eq:Schrodinger}) for helium and associated harmonics polarization (inset). Solid curves correspond to predictions based on RPO analysis (the two right curves correspond to closest return energy and the left curve corresponds to return within $5~{\rm a.u.}$, see text). The dashed curve corresponds to $2~{\rm Up}+{\rm Ip}$ and the rightmost solid curve overlaps with $3.17~{\rm Up}+{\rm Ip}$ harmonic radiation frequency (see text). In both panels the harmonic analysis is restricted to the duration of the plateau of the laser pulse (see text).}
\end{figure}

The interaction of a strong, short laser pulse with atoms/molecules is of great interest to the strong field and attosecond science communities because of the insights it provides in probing matter at the atomic scale~\cite{Beck08}.  These systems have been investigated from various angles ranging from experimental to numerical and analytical approaches using quantum, semi-classical and classical models (see Refs.~\cite{Beck08,Figu11,Beck12} and references therein). Among these, (semi)-classical models offer an insightful trajectory interpretation of the electronic dynamics which often compensates for the loss of purely quantum mechanical effects. A famous example is the recollision mechanism~\cite{Cork93,Scha93,Kuch87} which, for linearly polarized fields, explains many events such as nonsequential double (or multiple) ionization~\cite{Beck08}, high order harmonic generation (HHG)~\cite{Klin08,Kapt07} or laser induced electron diffraction~\cite{Zuo96,Meck08,Pete11}. The quantum nature of the system at hand raises the question of the applicability of such a classically-based interpretation. In this Letter we investigate the quantum-classical correspondence in the framework of HHG. Using quantum mechanical simulations, we demonstrate the existence of atomic HHG with an intense circularly polarized laser field and show that some of the harmonics are circularly polarized (see Fig.~\ref{fig:HHG_spectrum}). The properties of the HHG spectrum are later explained through specific classical electronic trajectories. The perspective offered by nonlinear dynamics allows one to fully interpret the observed HHG spectra and to devise quantitative predictions which are not accessible by standard interpretation~\cite{Frol09,Haes11}. The close connection we reveal between HHG and classical trajectories with circular polarization opens a way for controlling the highly nonlinear radiation spectrum properties through, e.g., nonlinear dynamical tools applied to orbit stability control~\cite{Huan07}. The robustness of the process with the initial preparation of the system, the atomic species and the laser parameters, along with the strong intensity of the radiation, hints at the universality of the mechanism described in this Letter and its accessibility to experimental observations with currently available intense laser technology.

Classical simulations have shown that not all two electron configurations lead to recollision in a strong, circularly polarized field~\cite{Maug10_3,Wang10,Kamo13}. It has been shown that the absence of recollision for tightly bound atoms is due to the unfavorable energy configuration of the system~\cite{Kamo13}: The ground state configuration yields ionized electron with energies too low to manage to recollide thereafter. On the other hand, for the same atomic systems, initializing the system with a higher energy, i.e., an excited initial condition, should compensate for the energy deficiency and make recollision available. From the quantum mechanical point of view, such an excited initial condition is obtained with an electronic wave packet
$\left|\psi\left(t=0\right)\right\rangle = \alpha_{0} \left|\psi_{0}\right\rangle + \sum{\alpha_{i} \left|\psi_{i}\right\rangle},$
where $\left|\psi_{0}\right\rangle$ is the ground state and $\left|\psi_{i}\right\rangle,\ i\in\mathbb{N}^{*}$ label excited states, $\alpha_{i}\in\mathbb{C}$ with $\sum_{i}{\left|\alpha_{i}\right|^{2}}=1$ for normalization. As we shall see later, the actual choice of the coefficients $\alpha_{i}$ is of little importance for observing atomic HHG providing the system initially contains a part of ground state and a part of excited state(s) $0<\left|\alpha_{0}\right|<1$. Yet it is worth noting that computed HHG spectra does not rely on coherence criteria for the preparation of the initial state. All figures reported in this Letter correspond to the arbitrary choice $\alpha_{0}=\alpha_{1}=1/\sqrt{2}$.

\begin{figure}
	\centering
	\includegraphics[width=\linewidth]{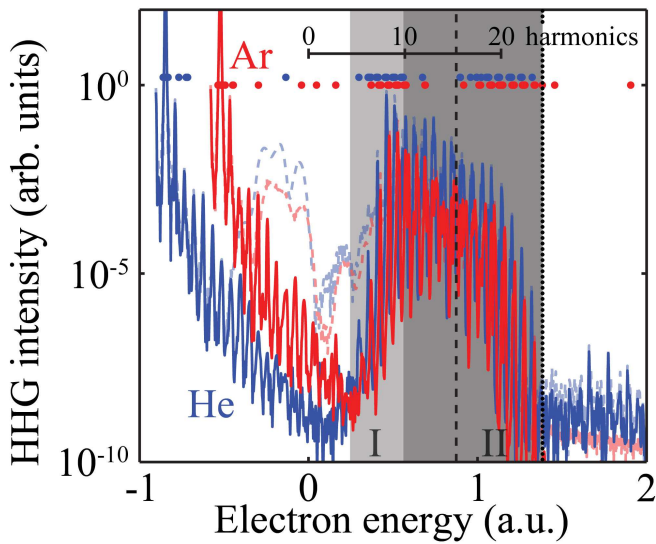}
	\caption{\label{fig:Return_energy} 
  (color online) Return electron energy $\mathcal{E}$ deduced from the harmonics spectrum for helium and argon (see labels). For each atom we compare the total pulse (dashed curves) and constant field (solid curves) spectra. For each spectrum, dots label harmonics for which the polarization is circular (larger than $0.9$, during the plateau). Gray areas correspond to return energy predictions provided by the RPO analysis. The vertical dashed and dotted lines label $2~{\rm Up}$ and $3.17~{\rm Up}$ electron energy respectively. We also display the harmonics scale. The computations are carried out at $4\times10^{14}~{\rm W}\cdot{\rm cm}^{-2}$ and~$800~{\rm nm}$.}
\end{figure}

Although the existence of recollision with circular polarization is now established~\cite{Maug10_3,Wang10,Kamo13}, it does not necessarily lead to HHG, and specifically circularly polarized HHG: Standard electric dipole transition selection rules advocate against it because the transition from a strongly excited electron to the ground state implies a variation of the magnetic quantum number by a large amount. Contrary to linear polarization, the instantaneous amplitude of a circular laser field is constant and never vanishes [see Eq.~(\ref{eq:Laser_field})]. As a consequence, for almost the entire duration of the pulse the Coulomb potential is strongly dressed by the laser electric field. A quick estimation shows that the energy of the first excited state of helium overcomes the field induced barrier for laser intensities larger than $10^{13}\ {\rm W}\cdot{\rm cm}^{-2}$, while the ground state energy does so for intensities larger than $3\times10^{15}\ {\rm W}\cdot{\rm cm}^{-2}$. It means that, in the context of strong field physics, atomic HHG with circular polarization corresponds to a transition/interaction from the continuum (over the barrier) to the ground state, rather than state to state. An interesting consequence of the elimination by over the barrier ionization of all excited states is a selection rule for the returning electron which can only recollide/interact with the ground state. The return energy of the electron $\mathcal{E}$ is deduced from the HHG spectrum using the relation
\begin{equation} \label{eq:Return_energy}
  \hbar \omega_{\rm HHG} = \mathcal{E} + {\rm Ip},
\end{equation}
where $\omega_{\rm HHG}=2\pi{\nu_{\rm HHG}}$ is the harmonic radiation frequency and ${\rm Ip}$ the ionization potential. From Eq.~(\ref{eq:Return_energy}), we deduce the electron return energy $\mathcal{E}=\hbar\omega_{\rm HHG}-{\rm Ip}$. This allows us to compare the return electron energy spectra for various atoms, such as helium and argon, as in Fig.~\ref{fig:Return_energy}.  For both atoms, we obtain a dominant peak (exceeding the height of the displayed box) in the lower part of the spectrum, which corresponds to the fundamental (laser) driving frequency. The peak is followed by a broad band of radiation (dashed curves) with negative energy ($\mathcal{E}$), and restricted to the ramp-up of the field, which corresponds to transitions from the dressed bound states to the ground state when the effective intensity of the laser is low. This part of the spectrum vanishes later on and reveals a gap in the electron return energy before a strong revival of the signal. We also notice that in specific regions of the spectra (denoted I and II), HHG tend to lock to circularly polarized emission. In what follows, we focus on the high harmonic part of the spectrum which is generated during the pulse plateau [solid curves, see Eq.~(\ref{eq:Laser_field})].

In order to simulate the electronic dynamics and compute the associated HHG spectrum, we solve the time dependent Schr\"{o}dinger equation numerically for a one active electron in two spatial dimensions (2D) given as follows (atomic units are used unless otherwise specified)
\begin{equation} \label{eq:Schrodinger}
	i \partial_{t} \psi\left({\bf x},t\right) = \left(-\frac{\Delta}{2} + V\left({\bf x}\right) + E\left({\bf x}, t\right)\right) \psi\left({\bf x}, t\right).
\end{equation}
We use a soft-Coulomb potential~\cite{Java88} with the softening parameter $a$ adjusted to model the atom under consideration~\cite{Note1}, corresponding to averaging over the dimension perpendicular to the field, $V\left({\bf x}\right)=-1/\sqrt{\left|{\bf x}\right|^{2}+a^{2}}$. The nucleus is assumed fixed at the origin. In the dipole approximation, the laser-matter interaction is given by
\begin{equation} \label{eq:Laser_field}
	E\left({\bf x},t\right) = \frac{E_{0}}{\sqrt{2}} f\left(t\right) \left(x \cos \omega t  + y \sin \omega t \right),
\end{equation}
where $E_{0}/\sqrt{2}$ is the peak field amplitude (corresponding to the intensity $I_{0}$) and $f$ corresponds to the envelope of the field. All figures reported in this Letter correspond to a trapezoidal envelope with a 2 laser cycle ramp-up, a 20 laser cycle plateau, and a 2 laser cycle ramp-down. The wavelength is $800~{\rm nm}$. We have checked that the results are robust with the pulse duration and wavelength. Radiation spectra are computed from the Fourier transform of the dipole acceleration \{${\bf R}_{\nu_{\rm HHG}} = \mathcal{F}\left[\left\langle \psi\left(t\right) | \ddot{\bf x}\left(t\right) | \psi\left(t\right) \right\rangle\right]\left(\nu_{\rm HHG}\right)$\}~\cite{Haes11}. The harmonics intensity is defined as the sum of the spectra in the $x$- and $y$- directions squared ($I_{\nu_{\rm HHG}} = \left|{\bf R}_{\nu_{\rm HHG}}\right|^{2}$), while the ellipticity accounts for the relative amplitude and phase between the two components. In order to avoid artifacts in the harmonics spectra, due to non-periodic temporal dipole acceleration signals, a Hanning window is used~\cite{MesPowSpec}.

\begin{figure}
	\centering
	\includegraphics[width=\linewidth]{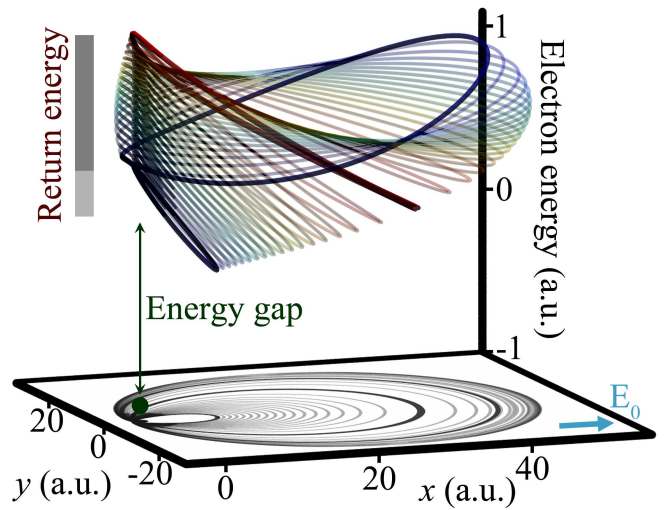}
	\caption{\label{fig:RPO_return_energy} 
  (color online) RPO family positions, in a frame rotating with the field ($\mathcal{O}_{2}$, see~\cite{Kamo13}), versus electron energy. Full-color trajectories correspond to the limiting orbits of the family (used for the radiation properties analysis). For all orbits, recolliding portions (closer than $5~{\rm a.u.}$ from the core) are displayed with a darker color. Accessible return energy for the electron is indicated with the gray stripes, using the same color code as in Fig.~\ref{fig:Return_energy}. We also display the direction of the laser field (arrow) and the ground state of helium (ball at the origin). Laser parameters are the same as in Fig.~\ref{fig:Return_energy}.}
\end{figure}

In Fig.~\ref{fig:HHG_spectrum}, HHG spectra are displayed for various laser intensities and confirm the characteristics of the spectra observed in Fig.~\ref{fig:Return_energy}. It shows that HHG is restricted to a narrow band of harmonics, with a gap of missing harmonics in the lower part of the spectra. The polarization of HHG radiation is random apart from two specific regions (I and II) where they tend to lock to circularly polarized emission. All of these properties are closely related to a set of periodic orbits, called recolliding periodic orbits (RPO)~\cite{Kamo13}, that organize the (classical) recollision dynamics with circular polarization. The close connection between the properties of the quantum HHG spectra and the properties of the RPOs lays the foundations of the quantum-classical correspondence reported in this Letter. In short, RPOs are classical periodic orbits observed in a frame rotating with the field~\cite{Kamo13}. They come in families and are composed of one or several loops that connect the core to ionized regions. They organize recollisions in the sense that a typical recolliding trajectory mimics RPOs in its journey back to the core. As an illustration, we display an RPO family in Fig.~\ref{fig:RPO_return_energy} as a function of position and energy of the electron. It has been shown that the determinant factor in RPO properties is the Coulomb tail of the potential ($-1/\left|{\bf x}\right|$) rather than its specific shape~\cite{Kamo13}, thus extending its influence on electron trajectories to large distances. In this perspective, it has been noticed that using a hard Coulomb potential ($a=0$) provides a very good description of the returning process irrespective of the atom. Already, this ``universal feature'' of RPOs predicts that the energy return [$\mathcal{E}$, see Eq.~(\ref{eq:Return_energy})] spectra for helium and argon should be qualitatively the same. This is indeed what is observed in Fig.~\ref{fig:Return_energy} where the electron energy spectra between $0.25$ and $1.4~{\rm a.u.}$ (gray regions) for helium and argon almost perfectly overlap. 

Circular polarization clearly attributes a well identified role to the laser and Coulomb potential: Generally speaking, the laser pulls the electron away~\cite{Cork11} while the Coulomb potential tends to recall it to the core. This interplay between the laser and Coulomb interactions, both playing an equally important role, is at the core of recollision in CP fields~\cite{Kamo13}. The importance of the Coulomb potential interaction in HHG has been noticed previously~\cite{Sand99,Host10,Shaf12}. Here, the potential role is revealed through the RPOs and it has to be accounted for in the return energy analysis: We define the classical energy of the electron as 
$\mathcal{E}=\left|{\bf p}\right|^{2}/2-1/\left|{\bf x}\right|,$
where ${\bf p}$ is the electron momentum. Recombination/interaction with the ground state is strongest for a returning electron at the closest point to the core. We scan through one of the RPO families (the one considered in Ref.~\cite{Kamo13}), record the energy at this location and compare it to the energy spectrum obtained with quantum simulations. The corresponding energy range is indicated by the dark gray region in Figs.~\ref{fig:Return_energy} and~\ref{fig:RPO_return_energy}. We see that it matches well the upper cut-off of the spectra of Fig.~\ref{fig:Return_energy} but it misses slightly lower harmonics. Since the ground state is not perfectly localized on the nucleus, an interaction is made possible within a small area around the core. Extending the possibility for recombination to $5~{\rm a.u.}$ from the nucleus (the size of the ground state for helium model) yields a larger range of possible return energies for the electron: It extends the lower return energy limit while the upper one is unchanged (see Fig.~\ref{fig:RPO_return_energy}), and corresponds to the light gray area in the figures. There, we see that both cut-offs are well predicted by the RPO analysis. 

In order to assess the robustness of the proposed mechanism, we investigate the prediction given by the RPO analysis as the laser intensity is varied. Following the RPOs with the laser parameters shows that, globally, they vary energetically like $2~{\rm Up}$. Looking at Figs.~\ref{fig:HHG_spectrum} and~\ref{fig:Return_energy}, we see that the $2~{\rm Up}$ rule of thumb (dashed curves) indeed provides an accurate overall guide for HHG and electron return energy spectra. In addition, we see in Fig.~\ref{fig:HHG_spectrum} that the prediction given by RPOs matches very well the radiation strip observed in HHG as intensity is varied (see continuous curves). The maximum return energy provided by the RPO analysis approaches $3.17~{\rm Up}$ (dotted vertical line in Fig.~\ref{fig:Return_energy}) as the laser intensity is increased, where ${\rm Up}=I_{0}/\left(4m_{e}\omega^{2}\right)$ is the ponderomotive energy. For linearly polarized fields, it is well known that $3.17~{\rm Up}$ is the maximum return energy for the electron~\cite{Cork93} based on the standard recollision picture. Here, the appearance of this number in the RPO analysis is unexpected since the standard recollision picture does not apply for circular polarization~\cite{Cork93,Cork11,Yuan12}. From the energy analysis of the RPOs, we also find a forbidden return energy range to the ground state (see arrow in Fig.~\ref{fig:RPO_return_energy}). This gap of forbidden return energy in the RPOs mirrors the gaps observed in HHG and electron energy spectra. A similar range of missing harmonics in the lower part of the spectrum followed by a restricted strip of strong revival of the HHG intensity has been reported for benzene molecule with circular polarization~\cite{Baer03}, which demonstrates the universality of the mechanism described here, beyond atomic systems.

With CP fields and atomic targets, recollisions and therefore HHG radiation is made possible in all directions. As a consequence, the polarization of emitted radiation is characteristic of the dynamics. Unstructured or unorganized recollisions would be expected to show-up randomly in time, leading to random amplitudes and phases in the $x$- and $y$-directions and ultimately random ellipticities. Circularly polarized HHG requires both the relative amplitudes to be equal and the phases to differ by $\pi/2$~\cite{Smir09,Yuan12}. Generally speaking, RPOs are (highly) unstable such that although they drive the overall recollision process, they do not manage to produce the long time organization required for CP radiation emission, with the exception of the upper and lower parts of the family which are stable (or weakly unstable). Looking at the polarization analysis of the radiation spectra for helium and argon displayed in Fig.~\ref{fig:Return_energy} (dots) we see that, with the exclusion of a few random return energies, CP harmonics are concentrated in the cut-off regions (I and II) of the radiation strip, which correspond to lower and upper extremes of the RPO family, where the orbits are stable or least unstable. This picture is confirmed by Fig.~\ref{fig:HHG_spectrum} (inset) as the laser intensity is varied.

Numerical simulations show that the results are very robust with the preparation of the initial (field-free) states, i.e., the symmetry of the excited states ($p$ or $s$), the phase between the ground and excited states and the relative weights between the two states. All the results converge to the observation that the intensity varies quadratically with the parameters $\alpha_{0,1}$ as 
$I_{\nu_{\rm HHG}} \propto \left|\alpha_{0} \alpha_{1}\right|^{2}$. 
The strongest spectrum is obtained for equal populations $\left|\alpha_{0}\right|=\left|\alpha_{1}\right|$, and more generally a strong radiation revival is observed roughly as long as $0.2\leq \left|\alpha_{0,1}\right|^{2}\leq 0.8$, thus implying an initial superposition of electronic states. The independence of the results with the phase between the states indicate that atomic HHG with circular polarization do not rely on any coherence criterion. This is explained by the fact that all excited states have an energy greater than the barrier and therefore vanish due to the field dressing: No matter which excited state(s) one starts from, all are coupled to the dressed continuum that forms a complex electronic cloud around the nucleus.


To summarize, the existence of atomic HHG, together with the specific properties of the spectra, strengthens the importance of recollision in circularly polarized laser fields. The quantum-classical correspondence between the HHG spectra and the properties of RPOs highlights the pivotal role of the Coulomb interaction in the recollision process. Through this correspondence and the properties of RPOs, we have fully interpreted the HHG spectra, and have shown that: $(1)$ Atomic HHG with circular polarization is restricted to a narrow band of harmonics and their intensity varies quadratically with the ground and excited state initial composition; $(2)$ the lower part of the spectra exhibits a gap of missing harmonics due to a forbidden range of electron return energies; $(3)$ in two specific regions, harmonics tend to lock to circularly polarized emission. The robustness of the process to laser parameters, target species and initial conditions should allow for experimental verification and extension to molecular systems and to generation of circularly polarized attosecond pulses~\cite{Yuan12}


The authors thank RQCHP and Compute Canada for access to massively parallel computer clusters and the CIPI for financial support in its ultrafast science program.
F.M.\ and A.D.B.\ acknowledge financial support from the Centre de Recherches Math\'ematiques. A.D.B.\ acknowledges financial support from the Canada Research Chair.
A.K.\ acknowledges financial support from the Chateaubriand fellowship program of the Embassy of France in the United States. 
A.K.\ and T.U.\ acknowledge funding from the NSF. 
The research leading to these results has received funding from the People Programme (Marie Curie Actions) of the European Union's Seventh Framework Programme FP7/2007-2013/ under REA grant agreement 294974. 



\end{document}